\documentclass[12pt,a4paper]{article}%
\usepackage{amsmath}%
\usepackage{amsfonts}%
\usepackage{amssymb}%
\usepackage{graphicx}
\usepackage{latexsym}
\usepackage{amsopn}
\usepackage{latexsym}
\usepackage[latin1]{inputenc}
\providecommand{\U}[1]{\protect\rule{.1in}{.1in}}
\newtheorem{theorem}{Theorem}
\newtheorem{acknowledgement}[theorem]{Acknowledgement}

\begin{document}
\clearpage

\begin{center}
Supersymmetrizing Massive Gravity
\end{center}
\vspace{0.5cm}
\begin{center}
O. Malaeb \\
Physics Department, American University of Beirut, Lebanon
\end{center}
\vspace{3cm}
%
%\title{Supersymmetrizing Massive Gravity}
%\author{O. Malaeb}
%\date{}
%\maketitle

\begin{abstract}
When four scalar fields with global Lorentz symmetry are coupled to gravity and take a vacuum expectation value breaking diffeomorphism invariance spontaneously, the graviton becomes massive. This model is supersymmetrized by considering four $N=1$ chiral superfields with global Lorentz symmetry. The global supersymmetry is promoted to a local one using the rules of tensor calculus of coupling the $N=1$ supergravity Lagrangian to the four chiral multiplets. When the scalar components of the chiral multiplets $z^A$ acquire a vacuum expectation value, both diffeomorphism invariance and local supersymmetry are broken spontaneously. The global Lorentz index A becomes identified with the space-time Lorentz index making the scalar fields $z^A$ vectors and the chiral spinors $\psi^A$ spin-$3/2$ Rarita-Schwinger fields. We show that the spectrum of the model in the broken phase consists of a massive spin-$2$ field, two massive spin-$3/2$ fields with different mass and a massive vector. 
\end{abstract}

\clearpage

\section{Introduction}
Massive gravity was the attraction of many authors. In 1970, van Dam, Veltman and Zakharov \cite{vandam}, \cite{zakharov} showed that there is a discrete difference between theories with zero mass graviton and theories with a small but non-zero mass. It was concluded that the graviton mass is is not some extreme small value and that it should be rigorously zero. They used the action of Fierz and Pauli \cite{fierzpauli} where general coordinate invariance is broken by mass terms. Their results don't go over into that of General Relativity in particular for the bending of light by the sun. Massive Gravity was first thought to be not physically possible because of this van Dam-Veltman-Zakharov discontinuity \cite{zakharov}. However, Vainshtein showed that the perturbation theory is not suitable when the mass goes to zero because of the singularity in the graviton mass in the higher orders contributions. Then, he resolved this problem \cite{vainshtein} by finding that the massive graviton behaves like a massless particle below a certain distance scale. So, the graviton could have a small mass without contradicting experiments. Further developments of this scale were considered in \cite{five} (see also \cite{six}). \newline
However, Boulware and Deser \cite{deser} investigated the behaviour of the massive Einstein theory and they concluded that it is ill-behaved since the ghost scalar does not decouple at the nonlinear level. They deduced that general relativity is an isolated theory. Isham, Salam and Strathdee \cite{iss} formulated a Lagrangian theory describing the mixing of the graviton with a massive $2^{+} f$ meson. Chamseddine, Salam and Strathdee \cite{css} generalized this by introducing the mixing terms through a spontaneous symmetry breaking mechanism. Dvali, Gabadadze, and Porrati \cite{dgp} considered theories with extra dimensions where they considered a five-dimensional model. Their theory seems to be free of ghosts when considered around a true background but only in decoupling limit. \newline
It was believed that since there is no Higgs mechanism that is free of ghosts and returns a mass for the graviton, it fails to obtain consistent massive general relativity free of ghosts in four dimensions. Siegel \cite{siegel} considered open-string field theories. He used four scalars to restore diffeomorphism invariance. However, studying his theory around a trivial background shows that it is not free of ghosts. Then , Arkani-Hamed, Georgi and Schwartz \cite{ahgs} applied this to massive gravity to introduce general coordinate invariance, but they didn't obtain a ghost free model. 't Hooft \cite{hooft} (see also \cite{kaku}) considered the use of four scalar fields breaking general coordinate invariance by their vacuum expectation value. These scalars give mass to the graviton where at the end a massive spin-2 boson and a massive scalar survive. However, in the unbroken phase, the scalar fields kinetic energies include a ghost. In the broken symmetry phase, the massive graviton does not have a Fierz-Pauli term, and the ghost state remains coupled. \newline
In \cite{simpleandelegant}, Chamseddine and Mukhanov used four scalars with global Lorentz symmetry. They showed how to form massive gravity by using Higgs mechanism. The graviton will get mass after the four scalar fields acquire non-zero expectation values as a consequence of spontaneous symmetry breaking. Three scalar degrees of freedom are absorbed by the graviton, while one remains coupled. The graviton then becomes massive, with Fierz-Pauli mass term, and thus having five degrees of freedom. The action is simply given by Einstein action plus the action of the four extra scalar particles. The resulting theory is ghost free below scales related to Vainshtein scales. In \cite{puzzle}, the limit of massive gravity, as the mass of the graviton goes to zero, was studied below the Vainshtein scale. It was shown that it goes smoothly to Einstein gravity.  In \cite{quadratic}, massive gravity is presented by a simplified reformulation where a simpler quadratic action is found. \newline
Ghost-free theories of massive gravity were proposed in four dimensions in \cite{rhamgabadadze} and \cite{rhametal}. In \cite{rhametal}, non-linear theories were constructed and it was shown that the Hamiltonian constraint which projects out the Boulware-Deser ghost is maintained up to the quartic order. In \cite{hassanrosen}, non-linear massive gravity models, for flat fiducial metric, were proposed and the absence of the Boulware-Deser ghost was proven (see also \cite{hassanetal} for curved fiducial metric). Recently, the characteristics of the ghost-free Wess-Zumino massive gravity model with five degrees of freedom were analyzed in \cite{deserwaldron} (see also \cite{deseretal}). It was shown that it admits superluminal shock wave solutions and accordingly acausal, where ironically this originates from the constraint that eliminates the Boulware-Deser ghost. 

\paragraph{}
In this paper, we generalize the Higgs mechanism used in the formulation of massive gravity to obtain a theory of massive supergravity. When massive gravity is supersymmetrized, the graviton and the gravitino both become massive due to the breakdown of diffeomorphism invariance. This theory is interesting because it will end up with a massive spin-$3/2$ particle in addition to the gravitino. Therefore, two spin-$3/2$ particles exist which is similar to what we have in $N=2$ supergravity. To write our globally supersymmetric action, we use superfields and write an action in superspace using D-terms and F-terms. Then, we use the rules of tensor calculus to promote global invariance to a local one. We'll couple the Supergravity Lagrangian \cite{chamsbook} to the chiral and vector multiplets by the rules given in \cite{cremmer} (see also \cite{chamsbook}) to get our final action that is restricted by certain conditions that are discussed below.
In section 2 we generalize the bosonic case and write down the possible D-terms and F-terms. In section 3, we use tensor calculus to couple to supergravity and find the action in the linearized approximation that is ghost free and returns Fierz-Pauli term for the vierbein. Section 4 is the conclusion. In appendix 1, we list all the possible D-type terms that can be used to form our action. Appendix 2 present the notation and convention used.

\section{Generalizing the Bosonic Case}

To generalize the bosonic case, we use a instead of four scalar fields a set of four chiral superfields $\Phi^{A}\left( x,\theta,\overline{\theta}\right)$ subject to the conditions%
\[
\overline{D}_{\overset{.}{\alpha}}\Phi^{A}\left(  x,\theta,\overline{\theta}\right)  =0,
\]
where $A=0,1,2,3$ is a global Lorentz index. These chiral superfields are given by
\begin{equation}
\Phi_A = \varphi_A + i(\theta \sigma^{\mu} \bar{\theta})\partial_{\mu} \varphi_A - \frac{1}{4} \theta\theta \bar{\theta}\bar{\theta} \partial_{\mu}\partial^{\mu} \varphi_A + \sqrt{2} \theta \psi_A - \frac{i}{\sqrt{2}} \theta \theta \left(\partial_{\mu} \psi_A \sigma^{\mu} \bar{\theta}\right) + \theta\theta F_A.
\end{equation}
In the bosonic case, the action is written in terms of an induced metric $H^{AB} = g^{\mu\nu}\partial_{\mu} \varphi^{A} \partial_{\nu} \varphi^{B}$, and it is found to be \cite{simpleandelegant}
\begin{align*}
S=-\frac{1}{2 \kappa^2}%
%TCIMACRO{\dint }%
%BeginExpansion
{\displaystyle\int}
%EndExpansion
d^{4}x\sqrt{-g}R  & +\frac{m^{2}}{8}\int d^{4}x\sqrt{-g}\left(  H^{2}%
-H_{\,B}^{A}H_{\,A}^{B}\ \right)  \\
& +3\left(  \frac{1}{16}H^{2}-1\right)  ^{2}.%
\end{align*}
where $\kappa^2 = 8 \pi G$. \\
There are many other actions, all of which agree at the second order level but differ at cubic or higher orders. Expanding
\[
\varphi^{A}=\left(  x^{A}+\chi^{A}\right)  \ ,\quad\text{ }g^{\mu\nu}=\eta^{\mu\nu}+h^{\mu\nu}\ .
\]
after defining%
\[
H^{AB}=\eta^{AB}+\bar{h}_{\,}^{AB}\
\]
we find that
\[
\ \bar{h}_{\,}^{AB}=h^{AB}+\ \left(  \partial^{A}\chi^{B}+\partial^{B}\chi^{A}\right)  \ +\cdots
\]
The action then takes the form
\[
S=-\frac{1}{2\ }%
%TCIMACRO{\dint }%
%BeginExpansion
{\displaystyle\int}
%EndExpansion
d^{4}x\sqrt{-g}R+\frac{m^{2}\ }{8}\int d^{4}x\sqrt{-g}\left[  \ \left( \bar{h}^{2} -\bar{h}_{B}^{A}\bar{h}_{A}^{B}\right)  + \cdots\right].
\]
The full action can be written in terms of $\bar{h}_{B}^{A}$ since the metric perturbations around Minkowski transform similar to the infinitesimal transformations which keep Einstein action invariant, $\tilde {x}^{A} = x^{A} +\xi^{A}$, with $\chi^A$ instead of $\xi^A$. Therefore, the action up to second order terms is given by
\begin{align*}
S  & =\frac{1}{2}%
%TCIMACRO{\dint }%
%BeginExpansion
{\displaystyle\int}
%EndExpansion
d^{4}x\left[  \bar{h}_{B}^{A,C}\bar{h}_{A,C}^{B} - 2\bar{h}_{C}^{A,C}\bar
{h}_{A,D}^{\,\,D} + 2\bar{h}_{C}^{A,C}\bar{h}_{,A}\right.  \\
& \left.  - \bar{h}_{,A}\bar{h}^{,A}-m^{2}\left(  \bar{h}_{B}^{A}\bar{h}%
_{A}^{B} - \bar{h}^{2}\right)  \right].
\end{align*}
The field $H$ is quadratic in the field $\varphi^A$, and thus the action is at least quartic in the fields $\varphi^{A}$. We note that the zero and linear terms are cancelled through the higher order term (in the above case it is quartic). It would be interesting to generalize to the supersymmetric case what was done in \cite{rhametal} using the quadratic formulation in \cite{quadratic}; however, this is much more complicated and such a formulation needs a superspace formulation of supergravity \cite{wess}.

\paragraph{}
To form our generalized induced metric, we start by writing a quartic interaction
\[
D_{\alpha}\Phi_{A}D_{\beta}\Phi_{B}\overline{D}^{\overset{.}{\alpha}}%
\Phi^{\ast C}\overline{D}^{\overset{.}{\beta}}\Phi^{\ast D}%
M_{\overset{.}{\alpha}\overset{.}{\beta}CD}^{\alpha\beta AB}%
\]
where $M_{\overset{.}{\alpha}\overset{.}{\beta}CD}^{\alpha\beta AB}$ is a multispinor constructed in such a way as to make the action invariant under Lorentz transformations. There are two possible strategies to adopt: to symmetrize and antisymmetrize with respect to the fermionic indices $\alpha\beta$ and $\overset{.}{\alpha}\overset{.}{\beta},$ or to use the equivalence of $\alpha\overset{.}{\alpha}$ to a vector index%
\[
V_{\alpha\overset{.}{\alpha}}=\sigma_{\alpha\overset{.}{\alpha}}^{\mu}V_{\mu}.%
\]
We thus define $H_{ABC}$ as the basic field%
\begin{equation}
H_{ABC} = D^{\alpha} \Phi_A (\sigma_B)_{\alpha \dot{\alpha}} \bar{D}^{\dot{\alpha}} \Phi_C^{*} = D\Phi_A \sigma_B \bar{D} \Phi_C^*
\end{equation}
Its Hermitian conjugate is
\[
H_{ABC}^{\ast}=D\Phi_{C}\sigma_{B}\overline{D}\Phi_{A}^{\ast}=H_{CBA}%
\]
We also denote $H_{ABC} \eta^{AB}$ by $H_{AAC}$ and we define the contracted field
\begin{equation}
H_{C}=H_{AAC},\qquad H_{A}^{\ast}=H_{ACC}
\end{equation}
to simplify our expressions. The products that could be formed from this H field are given in appendix \ref{app1}. \newline
We can therefore start from the action
\begin{align*}
& c_{1}H_{ABC}H_{ABC}+c_{2}H_{ABC}H_{ACB}+c_{2}^{\ast}H_{ABC}H_{BAC} +c_{3}H_{ABC}H_{BCA} \\ 
& +c_{4}H_{ABC}H_{CAB}  +c_{5}H_{ABC}H_{CBA} + c_{6}H_{A}H_{A}^{\ast} + c_{7}H_{A}H_{A} + c_{7}^{\ast}H_{A}^{\ast}H_{A}^{\ast} \\
& + \epsilon^{ABCD}H_{ABC}\left(  c_{8}H_{D}-c_{8}^{\ast}H_{D}^{\ast}\right) +\epsilon^{ABCD}\left(  c_{9}H_{ABE}H_{CDE}+c_{9}^{\ast}H_{EAB} H_{ECD} \right. \\
& + \left. c_{10}H_{AEB}H_{CED}  + c_{11}H_{AEB}H_{ECD}+c_{11}^{\ast}H_{AEB}H_{CDE}
+c_{12}H_{EAB}H_{CDE}\right). 
\end{align*}
In addition to the D-type terms, we can add to our action F-type terms such as%
\begin{align}
{}& c_{13}\overline{D}^{2}\left(  D\Phi_{A}\sigma^{AB}D\Phi_{B}\right)
+c_{13}^{\ast}D^{2}\left(  \overline{D}\Phi_{A}^{\ast}\overline{\sigma}%
^{AB}\overline{D}\Phi_{B}^{\ast}\right) \nonumber \\
&  +c_{14}\overline{D}^{2}\left(D\Phi_{A}D\Phi^{A} \bar{D}\Phi_{B}^{*} \bar{D}\Phi^{B*} \right) 
 +c_{14}^{\ast}D^{2}\left(  \bar{D}\Phi_{A}^{*} \bar{D}\Phi^{A*} D\Phi_{B} D\Phi^{B} \right)
\end{align}
where all the constants $c_{i}$ are real except for those whose conjugate
appear (i.e. $c_{2},$ $c_{7},$ $c_{8},$ $c_{9},$ $c_{11},$ $c_{13},$ $c_{14}$
are complex). Much more F-type terms can be written, but we wrote only those we are going to use. 

\paragraph{}
In the bosonic case \cite{simpleandelegant} we have seen that the action with the correct behaviour is
expressed in terms of the field $\overline{h}_{AB}$ where
\[
\overline{h}_{AB} = H_{AB}-\eta_{AB} \equiv H_{AB} - \partial_{\mu} x_A \partial^{\mu} x_B %
\]
so that in this case there is no need to consider higher order terms in $H_{AB}$ and is enough to consider the terms
\[
\left(  \bar{h}_{B}^{A}\bar{h}_{A}^{B}-\bar{h}^{2}\right)
\]
For this, we consider instead to work with 
\[
\overline{H}_{ABC}=H_{ABC}-Dx_{A}\sigma_{B}\overline{D}x_{C}^{\ast}%
\]
where $x_A$ are the coordinates, since this will avoid including higher order terms in $H_{ABC}$. 
\allowdisplaybreaks 
As we'll prove below, at the end the action will be formed of three D-type terms and two F-type terms. It will be given by
\begin{align}
{}& m^4 \int \left(c_1 \bar{H}_{ABC} \bar{H}_{BCA} + c_2 \bar{H}_{ABB} \bar{H}_{CCA} + c_3 \bar{H}_{AB} \bar{H}_{AB}^*  \right)  d\theta^2 d\bar{\theta}^2 d^4x \nonumber \\
& + \frac{m^2}{\kappa} \int \left( c_4 \bar{D}^2\left(D\Phi_A \sigma^{AB} D\Phi_B\right) + c^{*}_4 {D}^2\left(\bar{D}\Phi^{*}_A \bar{\sigma}^{AB} \bar{D}\Phi_B^{*}\right) \right) d\theta^2 d^4x  \nonumber \\
& + m^4 \int c_5 \bar{D}^2\left(D\Phi_A D\Phi^A \bar{D} \Phi_B^{*} \bar{D} \Phi^{B *}\right) d\theta^2 d^4x \nonumber \\
& +m^4 \int  c_5^{*} D^{2}\left( \bar{D}\Phi_{A}^{*} \bar{D}\Phi^{A*} D\Phi_{B} D\Phi^{B} \right) d\theta^2 d^4x
\end{align}
where $H_{AB} = D\Phi_A  {D}\Phi_B$ and $m$ and $\kappa$ are used to fix the dimensions.

\section{Coupling to Supergravity}

To couple our supersymmetric action to supergravity, we first start by writing down the Supergravity Lagrangian. We need first to define the vierbein
\begin{equation}
e^{\mu}_a = g^{\mu \nu} e_{\nu a} = g^{\mu \nu} \eta_{ab} e^b_{\nu},
\end{equation}
and its relation to the spin connection $ w_{\nu a}^{\ \ b} $ is given by the equation
\begin{equation}
\partial_{\nu}{e^{\mu}_a} = - w_{\nu a}^{\ \ b} {e^{\mu}_b} - \Gamma^{\mu}_{\gamma \nu} e^{\gamma}_a
\end{equation}
The Supergravity Lagrangian field content consists of the spin 2 field, $e_{a \mu}$, the spin-$3/2$ field, $\psi_{\mu}$, and the auxiliary fields $S$, $P$, $A_{\mu}$. This Lagrangian is given by \cite{chamsbook}
\begin{equation}
L_{S.G} = -\frac{e}{2 \kappa^2} R(e,w) - \frac{e}{3} \left|u\right|^2 + \frac{e}{3} A_{\mu}A^{\mu} - \frac{1}{2} \bar{\phi}_{\mu} R^{\mu}
\end{equation}
where
\begin{align}
u = {}& S - iP \\
R_{\mu \nu}\, ^{rs} =& \partial_{\mu} w_{\nu}^{\ rs} + w_{\mu}^{\ rp} w_{\nu p}^{\ \ s} - \mu \leftrightarrow \nu \\
R^{\mu} =& \epsilon^{\mu \nu \rho \sigma} \gamma_{\nu} \gamma_{5} D_{\rho}(w) \phi_{\sigma} \\
R =& e_{r}^{\ \mu} e_{s}^{\ \nu} R_{\mu \nu}\, ^{rs} \\
D_{\mu} =& \partial_{\mu} + (1/2) w_{\mu rs} \sigma^{rs} \\
w_{\mu rs} =& w_{\mu rs}(e) + K_{\mu rs} (e, \phi_{\mu}) \\
K_{\mu rs}(e, \phi_{\mu}) =& (\kappa^2/4)(\bar{\phi}_{\mu} \gamma_r \phi_s - \bar{\phi}_{\mu} \gamma_s \phi_r + \bar{\phi}_{r} \gamma_{\mu} \phi_s )
\end{align}
and $e$ is the determinant of the vierbein. This lagrangian is invariant under local supersymmetry transformations up to a total divergence.

\paragraph{}
Next we couple this to the supersymmetric action using the rules of tensor calculus. These rules are known and they provide us by the method of coupling Supergravity to the components of vector and chiral multiplets \cite{chamsbook} (see also \cite{cremmer}). The global supersymmetry will then be promoted to a local one. Below is a review of how this is done.\\
The component fields of a vector multiplet are Majorana spinors ($\xi$ and $\lambda$), two scalars ($C$ and $M$), and one auxiliary scalar field ($D$). It is given by  
\begin{equation}
V = \left(C, \xi, M, V_{\mu}, \lambda, D\right).
\end{equation}
A left-handed chiral multiplet (F-type) contains a complex scalar field $z$, left-handed Weyl spinors $X_L$, and a complex auxiliary field $h$. Then it is
\begin{equation}
F = \left(z, X_L, h\right).
\end{equation}
For the F-type multiplet the action formula is
\begin{equation}
e^{-1}L_{F} = h  + \kappa uz +  \kappa \bar{\phi}_{\mu} \gamma^{\mu} \chi + i \kappa^2 \bar{\phi}_{\mu} \gamma^{\mu \nu} \phi_{vR} z + h.c.
\end{equation}
while for the D-type multiplet one has
\begin{align}
e^{-1}L_{D} ={}& D + \frac{i \kappa}{2} \bar{\phi}_{\mu} \gamma^5 \gamma^{\mu} \lambda - \frac{\kappa}{3} \left(uM^* + u^* M\right)  + \frac{i \kappa^2}{8} \epsilon^{\mu \nu \rho \sigma} \bar{\phi}_{\mu} \gamma_{\nu} \phi_{\rho} \bar{\xi} \phi_{\sigma} \nonumber \\
&  + \frac{2}{3} \kappa V_{\mu} \left( A^{\mu} + \frac{3}{8} i e^{-1} \epsilon^{\mu \rho \sigma \tau} \bar{\phi}_{\rho} \gamma_{\tau} \phi_{\sigma} \right) - i\frac{\kappa}{3} e^{-1} \bar{\xi} \gamma_5 \gamma_{\mu} R^{\mu} \nonumber \\
&- \frac{2}{3} \kappa^2 C e^{-1} L_{S.G.} + e^{-1}L_{S.G.} 
\end{align}
After eliminating the auxiliary fields by their equations of motion and neglecting the terms that will later lead to  higher than quadratic orders, we get
\begin{align}
e^{-1}L_{F} + e^{-1}L_{D} ={}& D - \frac{1}{2 \kappa^2} R(e,w) - \frac{1}{2} e^{-1} \bar{\phi}_{\mu} R^{\mu} + h + h^* - \kappa^2 \left( Mz + M^{*}z^{*}\right)  \nonumber \\
&  + 3 \kappa^2 zz^{*} + \left(\kappa \bar{\phi}_{\mu} \gamma^{\mu} \chi + i \kappa^2 \bar{\phi}_{\mu} \gamma^{\mu \nu} \phi_{vR} z + h.c.\right).
\end{align}

\noindent Therefore, to write down the full Lagrangian, we have first to express the supermultiplets in terms of their component fields. Substituting for the metric $g^{\mu \nu} = e^{\mu}_a e^{\nu a}$ and expanding the fields around the vacuum solution
\begin{equation}
\varphi^A = x^A + \chi^A, \quad \quad e^{\mu}_{a} = \delta^{\mu}_{a} + \bar{e}^{\mu}_{a},
\end{equation}
then the components of our superfields, ignoring terms higher than quadratic order, are given by

\begin{enumerate}
\item For the superfield $\bar{H}_{ABC}\bar{H}^{BCA}$:
\begin{align}
{}& C = 0, \, \xi = 0, \, M = -8 \ \left(\psi_A \sigma_B \bar{\sigma}^A \psi^B \right) \nonumber \\
& V_{\mu} = \ \textnormal{quadratic}, \, \lambda = \ \textnormal{quadratic} \nonumber \\
& D = -16 \ \left(\partial_{\mu} \chi_A \partial^{\mu} \chi^A + \partial_{\mu} \chi_A^* \partial^{\mu} \chi^{A *} \right) + 32 \ \left( \partial_{\mu} \chi_A \partial^{\mu} \chi^{A *} + \partial^{A} \chi_A \partial^{B} \chi_B^* \right) \nonumber \\
& + 80 \ F_A F^{A *} - 8 \ \epsilon^{ABCD} \psi_A \sigma_B \partial_C \bar{\psi}_D - 8i \ \left(\psi_A \sigma^A \partial_B \bar{\psi}^B + \psi_A \sigma^B \partial^A \bar{\psi}_B\right) \nonumber \\
&-56 i\ \psi_A \sigma^{\mu} \partial_{\mu} \bar{\psi}^A + 32 \ \bar{e} \partial_A \chi^A + 32 \ \bar{e} \partial_A \chi^{A *} + 32 \ \bar{e}^2
\end{align}
where $V_{\mu}$ and $\lambda$ don't affect our results since they will give terms with higher orders. Similarly, the components of the other vector multiplets are calculated. Below we list only the two supermultiplets that at the end will enter in the action.
\item $\bar{H}_{AB}^{\quad B}\bar{H}_C^{\ CA}$: 
\begin{align} 
{}&  C = 0, \, \xi = 0, \, M = 16 \ \left(\psi_A \sigma^A \bar{\sigma}^B \psi_B \right) \nonumber \\
& V_{\mu} = \ \textnormal{quadratic}, \, \lambda = \ \textnormal{quadratic} \nonumber \\
& D = 32 \ \left(\partial_{\mu} \chi_A \partial^{\mu} \chi^A + \partial_{\mu} \chi_A^* \partial^{\mu} \chi^{A *} \right) + 32 \ \partial^{A} \chi_A \partial^{B} \chi_B^* + 128 \ \partial_{\mu} \chi_A \partial^{\mu} \chi^{A *} \nonumber \\
& + 272 \ F_A F^{A *} + 8 \ \epsilon^{ABCD} \psi_A \sigma_B \partial_C \bar{\psi}_D - 8i \ \left(\psi_A \sigma^A \partial_B \bar{\psi}^B + \psi_A \sigma^B \partial^A \bar{\psi}_B\right) \nonumber \\
& - 200 i \ \psi_A \sigma^{\mu} \partial_{\mu} \bar{\psi}^A + 96 \ \bar{e}^2 + 128 \ \bar{e}^a_{\mu} \bar{e}^{\mu}_a + 96 \ \bar{e} \partial_A \chi^A + 96 \ \bar{e} \partial_A \chi^{A *} \nonumber \\
& + 128 \ \bar{e}^{\mu}_A \partial_{\mu} \chi^A + 128 \ \bar{e}^{\mu}_A \partial_{\mu} \chi^{A *}
\end{align}
and the third 
\item $\left(D\Phi_A  {D}\Phi_B\right) \left(\bar{D}\Phi^{B *} \bar{D} \Phi^{A *}\right)$: 
\begin{align}
{}& C = 0 , \, \xi = 0, \, M = 0 \nonumber \\
& V_{\mu} = 0, \, \lambda = \ \textnormal{quadratic} \nonumber \\
& D = 32 \ \left( \partial^{A} \chi_A \partial^{B} \chi_B^* + \partial_{\mu} \chi_A \partial^{\mu} \chi^{A *} \right) + 80 \ F_A F^{A *} - 8 \ \epsilon^{ABCD} \psi_A \sigma_B \partial_C \bar{\psi}_D \nonumber \\
& - 8i \ \left(\psi_A \sigma^A \partial_B \bar{\psi}^B + \psi_A \sigma^B \partial^A \bar{\psi}_B\right) - 56 i \ \psi_A \sigma^{\mu} \partial_{\mu} \bar{\psi}^A + 64 \ \bar{e}^a_{\mu} \bar{e}^{\mu}_a \nonumber \\
& + 64 \ \bar{e}^{\mu}_A \partial_{\mu} \chi^A + 64 \ \bar{e}^{\mu}_A \partial_{\mu} \chi^{A *}
\end{align}
\end{enumerate}

From the above vector multiplets, we can form an action with the following required conditions
\begin{itemize}
\item It has a Fierz-Pauli term for the vierbeins $(\bar{e}^{\mu}_A \bar{e}^A_{\mu} - \bar{e}^2)$
\item It contains no linear vierbein term
\item It gives Maxwell form for the $\chi_A$ fields 
\begin{equation}
l \left( \partial_{\mu}\chi_A \partial^{\mu} \chi^{A*} - \partial_A \chi^A \partial_B \chi^{B*}\right) 
\end{equation}
where $l$ is a constant
\item It is ghost free where there should be no terms like
\begin{equation} 
\partial_{\mu}\chi_A\partial^{\mu}\chi^A, \quad \text{or} \quad \partial_A \chi^A \partial_B \chi^B
\end{equation}
\item The gravitinos should be massive
\end{itemize}
The first four conditions are well satisfied if we only consider D-type terms. However, to make the gravitino massive, F-type terms should be included. It is found out that only such terms will return a mass term for the gravitino. 
\paragraph{}
Several F-type terms can be written. Calculations show that the two terms listed above will give us the required results. Their components are given by
\begin{enumerate}
\item $\bar{D}^2 \left(D\Phi_A \sigma^{AB} D\Phi_B \right)$:
\begin{align}
{}& z = -96 i -48 i \ \partial^{A} \chi_A - 48 i \ \bar{e} - 16 i \ \bar{e} \partial^A \chi_A   - 32i \ \bar{e}^{\mu}_A \partial_{\mu} \chi^A + 8i \ \bar{e}_{\mu}^A \bar{e}^{\mu}_A \nonumber \\
&  \quad \quad - 8i \ \bar{e}^2 \nonumber \\
& X_{\alpha} = -4\sqrt{2} \ \left(\sigma^{AB} \psi_A\right)_{\alpha} \partial_{\mu} \partial^{\mu} \chi_B - 16\sqrt{2} i \ \left(\partial^A \psi_A\right)_{\alpha} \partial^B \chi_B  \nonumber \\
& \quad \quad - 48 \sqrt{2} i \ \left(\partial^A \psi_A\right)_{\alpha} + 16\sqrt{2} i \ (\partial^A \psi_B)_{\alpha} \partial^B \chi_A \nonumber \\
& h = \ \textnormal{total derivative}
\end{align}
$h$ is a total derivative, therefore it won't affect our calculations. The components of the other term are 
\item $\bar{D}^2 \left(D\Phi_A D\Phi^A \bar{D}\Phi_B^{*} \bar{D}\Phi^{B *} \right):$ 
\begin{align}
{}& z = -128\ \bar{\psi}_A \bar{\psi}^A \nonumber \\
& X_{\alpha} = 128 \sqrt{2} i \ \left(\sigma^{B} \bar{\psi}_B\right)_{\alpha} \partial^{A} \chi_A + 64 \times 4 \sqrt{2} i \ \left(\sigma^{\nu} \bar{\psi}_B\right)_{\alpha} \partial_{\nu} \chi_B^{*}  \nonumber \\
& h = -64\times 4 \ \left( \partial_{\mu} \chi_A \partial^{\mu} \chi^{A} + \partial_{\mu} \chi_A^* \partial^{\mu} \chi^{A *} + \partial^{A} \chi_A \partial^{B} \chi_B^* \right) - 64\times 2 \ \ F_A F^{A *} \nonumber \\
& - 64 \ \epsilon^{ABCD} \psi_A \sigma_B \partial_C \bar{\psi}_D + 64 i \ \left(\psi_A \sigma^A \partial_B \bar{\psi}^B + \psi_A \sigma^B \partial^A \bar{\psi}_B + \psi_A \sigma^{\mu} \partial_{\mu} \bar{\psi}^A \right) \nonumber \\
& -64 \times 16 - 64\times 16 \ \bar{e} - 64\times 8 \ \bar{e}^A_{\mu} \bar{e}^{\mu}_A -64\times 4 \ \bar{e}^2  \nonumber \\
& + 64 \ \left(\bar{e} \partial_{A} \chi^A + \bar{e} \partial_{A} \chi^{A *}\right) + 64\times 4 \ \left(\bar{e}^{\mu}_A \partial_{\mu} \chi^A + \bar{e}^{\mu}_A \partial_{\mu} \chi^{A *}\right).
\end{align}
\end{enumerate}

Forcing the constraints mentioned above to obtain a well behaved action, we can write a system of equations to solve for the constants $c_1, c_2, c_3, c_4$ and $c_5$. The equations are found to be 
\begin{itemize}
\item No ghost: $-16 c_1 + 32 c_2 - 64 \times 4 \left( c_5 + c_5^* \right) = 0$
\item Maxwell: $32 c_1 + 64 \times 2 c_2 + 32 c_3 = l$ \\
 and $32 c_1 + 32 c_2 + 32 c_3 + 48^2 \times 3 \ c_4 c_4^* - 64 \times 4 \left( c_5 + c_5^* \right) = - l$
\item Constant: $-64 \times 16 \left(c_5 + c_5^* \right) + 96^2 \times 3 \ c_4 c_4^* = 0$
\end{itemize}
The solution is 
\begin{align}
c_1 ={}& \frac{l}{24} - 432 \ c_4 c_4^* \nonumber \\
c_2 ={}& \frac{l}{48} \nonumber \\
c_3 ={}& -\frac{3l}{32} + 432 \ c_4 c_4^* \nonumber \\
c_5 ={}& \frac{27}{2} \ c_4 c_4^*
\end{align}
By normalizability of the kinetic term of the $\chi^A$ field, we have $l = -1/2$. Moreover, $c_4$ is arbitrary, but we can choose it such that the term $\bar{\psi}_A \gamma^A \gamma^B \psi_B$ cancels out. This sets $c_4$ to be $\frac{i}{48 \sqrt{6}}$, and reduces the full Lagrangian $(e^{-1}L_{F} + e^{-1}L_{D})$ to
\begin{align*}
{}& -\frac{1}{2} m^4 \ \left(\partial_{\mu} \chi_A \partial^{\mu} \chi^{A*} - \partial^{A} \chi_A \partial^{B} \chi_B^*\right) + \frac{7}{3} m^4 \ \left(\bar{e}^A_{\mu} \bar{e}^{\mu}_A - \bar{e}^2\right) - m^4 \ F_A F^{A *} \nonumber \\
& -\frac{7}{3} m^4 \ \left(\bar{e} \partial_A \chi^A +  \bar{e} \partial_A \chi^{A *}\right) + \frac{7}{3} m^4 \ \left(\bar{e}^{\mu}_A \partial_{\mu} \chi^A + \bar{e}^{\mu}_A \partial_{\mu} \chi^{A *}\right) \nonumber \\
& -\frac{5}{24} m^4 \ \epsilon^{ABCD} \bar{\psi}_A \gamma_B \gamma_5 \partial_C {\psi}_D + \frac{3i}{8} m^4 \ \bar{\psi}_A \gamma_{\mu} \partial^{\mu} {\psi}^A - \frac{\sqrt{6}}{8} m^6 \kappa \ \bar{\psi}_A \psi^A \nonumber \\
& + \frac{\sqrt{6}}{18} m^6 \kappa \ \bar{\psi}_A \gamma^A \gamma^B \psi_B - \frac{5 \sqrt{6}}{36} m^6 \kappa \ \bar{\psi}_A \gamma^B \gamma^A \psi_B + \frac{1}{2} e^{-1} \epsilon^{\mu \nu \rho \sigma} \bar{\phi}_{\mu} \gamma_5 \gamma_{\nu} \partial_{\rho} \phi_{\sigma} \nonumber \\
& + \frac{\sqrt{2} i}{4} m^4 \kappa \ \bar{\phi}_{\mu} \gamma^{\mu} \gamma^A \psi_A -  \frac{\sqrt{2} i}{4} m^4 \kappa \ \bar{\psi}_A \gamma^A \gamma^{\mu} \phi_{\mu} + \frac{\sqrt{3}}{6} m^2 \bar{\phi}_{\mu} \gamma^{\mu} \partial^A \psi_A  \nonumber \\
& + \frac{\sqrt{3}}{6} m^2 \partial^A \bar{\psi}_A \gamma^{\mu} {\phi}_{\mu} + \frac{\sqrt{6} i}{3} m^2 \kappa\ \bar{\phi}_{\mu} \gamma^{\mu \nu} \phi_{v} + \frac{\sqrt{3}}{12} m^2 \kappa \ \bar{\phi}_{\mu} \gamma^{\mu} \gamma^A \gamma^B \partial_B \psi_A \nonumber \\
& + \frac{\sqrt{3}}{12} m^2 \kappa \ \partial_B \bar{\psi}_A \gamma^{B} \gamma^A \gamma^{\mu} {\phi}_{\mu}  - \frac{1}{2 \kappa^2} R(e,w).
\end{align*}
Now it is clear how $m$ and $\kappa$ fix the dimensions, where we have $[\chi_A] = -1 [\bar{e}]=0, [F_A]=0, [\psi_A] = -1/2$ and the gravitino $[\phi_{\mu}] = 3/2$.

\paragraph{}
\noindent The equations of motion for $\bar{{\psi}}_A$ and $\bar{{\phi}}_{\mu}$ are respectively
\begin{align}
{}& \frac{-5}{24} m^4 \epsilon^{ABCD} \gamma_B \gamma_5 \partial_C {\psi}_D +  \frac{3i}{8} m^4 \gamma_{\mu} \partial^{\mu} {\psi}^A - \frac{ \sqrt{6}}{8} m^6 \kappa \ {\psi}^A - \frac{5 \sqrt{6}}{36} m^6 \kappa \ \gamma^B \gamma^A \psi_B \nonumber \\
& + \frac{\sqrt{6}}{18} m^6 \kappa \ \gamma^A \gamma^B \psi_B - \frac{\sqrt{3}}{6} m^2 \ \gamma^{\mu} \partial^A \phi_{\mu} - \frac{\sqrt{2}i}{4} m^4 \kappa \ \gamma^A \gamma^{\mu} \phi_{\mu} \nonumber \\
& - \frac{\sqrt{3}}{12} m^2 \kappa \ \gamma^B \gamma^A \gamma^{\mu} \partial_B \phi_{\mu} =0
\end{align}
and
\begin{align}
{}& \frac{\sqrt{3}}{6} m^2 \ \gamma^{\mu} \partial_A {\psi}^{A} + \frac{\sqrt{2}i}{4} m^4 \kappa \ \gamma^{\mu} \gamma^{\nu} \psi_{\nu} + \frac{\sqrt{3}}{12} m^2 \ \gamma^{\mu} \gamma^A \gamma^B \partial_B \psi_A + \frac{\sqrt{6}i}{3} m^2 \kappa \gamma^{\mu \nu} \phi_{\nu} \nonumber \\
& + \frac{1}{2} \epsilon^{\mu \nu \rho \sigma} \gamma_5 \gamma_{\nu} \partial_{\rho} {\phi}_{\sigma} = 0.
\end{align}
Next we can decompose $\psi_A$ into a spin-$3/2$ helicity, $\hat{\psi}_{A}$, and a spin-$1/2$ helicity, $\lambda$:
\begin{align}
{}& \psi_{A} = \hat{\psi}_{A} + \frac{1}{4} \gamma_A \gamma_5 \lambda \nonumber \\
&\Rightarrow \bar{\psi}_{A} = \bar{\hat{\psi}}_{A} + \frac{1}{4} \bar{\lambda} \gamma_A \gamma_5,
\end{align}
where $\gamma_A \hat{\psi}^A = 0$. Similarly, we decompose $\phi_{\mu}$
\begin{align}
{}& \phi_{\mu} = \hat{\phi}_{\mu} + \frac{1}{4} \gamma_{\mu} \gamma_5 \eta \nonumber \\
&\Rightarrow \bar{\phi}_{\mu} = \bar{\hat{\phi}}_{\mu} + \frac{1}{4} \bar{\eta} \gamma_{\mu} \gamma_5,
\end{align}
where $\gamma_{\mu} \hat{\phi}^{\mu} = 0$ and again $\hat{\phi}_{\mu}$ is a spin-$3/2$ helicity, while $\eta$ is spin-$1/2$ helicity. Using this decomposition, the equations of motion become
\begin{align}
{}& \frac{-5}{24} m^4 \epsilon^{ABCD} \gamma_B \gamma_5 \partial_C {\hat{\psi}}_D + \frac{5}{96} m^4 \epsilon^{ABCD} \gamma_B \gamma_D \partial_C {\lambda} +  \frac{3i}{8} m^4 \gamma_{\mu} \partial^{\mu} \hat{\psi}^A  \nonumber \\ 
& + \frac{3i}{32} m^4 \gamma_{\mu} \gamma^A \gamma^5 \partial^{\mu} \lambda + \frac{3 \sqrt{6}}{32} m^6 \kappa \ \gamma^A \gamma^5 \lambda - \frac{29 \sqrt{6}}{72} m^6 \kappa \ \hat{\psi}^A - \frac{\sqrt{3}}{6} m^2 \ \gamma^5 \partial^A \eta \nonumber \\
&  - \frac{\sqrt{2} i}{4} m^4 \kappa \ \gamma^{A} \gamma^5 \eta - \frac{\sqrt{3}}{12} m^2 \kappa \ \gamma^B \gamma^{A} \gamma^5 \partial_B \eta =0
\label{psi}
\end{align}
and
\begin{align}
{}& \frac{\sqrt{3}}{3} m^2 \ \gamma^{\mu} \partial_A \hat{{\psi}}^{A} + \frac{\sqrt{2}i}{4} m^4 \kappa \ \gamma^{\mu} \gamma^{5} \lambda + \frac{\sqrt{6}}{6} m^2 \kappa \ \hat{\phi}^{\mu} - \frac{\sqrt{6}}{8} m^2 \kappa \gamma^{\mu} \gamma^5 \eta  \nonumber \\ 
& + \frac{1}{2} \epsilon^{\mu \nu \rho \sigma} \gamma_5 \gamma_{\nu} \partial_{\rho} \hat{{\phi}}_{\sigma} + \frac{1}{2} \gamma^5 \gamma^{\mu \rho} \partial_{\rho} \eta= 0.
\label{phi}
\end{align}
To simplify these field equations, we multiply equations (\ref{psi}) and (\ref{phi}) by $\gamma_{A}$ and $\gamma_{\mu}$ respectively. 
%Then we get
%\begin{align}
%\frac{i}{3} \partial_A \hat{\psi}^A - \frac{i}{8} \gamma^5 \gamma^A \partial_A \lambda + \frac{3\sqrt{6}}{8} m^2 \kappa \ \gamma^5 \lambda - \sqrt{2}i \kappa \ \gamma^5 \eta = 0
%\label{psigamma}
%\end{align}
%\begin{align}
%\frac{i}{3} m^2 \ \partial_A \hat{\psi}^A - \frac{\sqrt{6}}{12} m^4 \kappa \ \gamma^5 \lambda - \frac{\sqrt{2}}{8} i m^2 \kappa \ \gamma^5 \eta + \frac{\sqrt{3}}{12} \partial_{\mu} \hat{\phi}^{\mu} + \frac{\sqrt{3}}{16} \gamma^5 \gamma^{\mu} \partial_{\mu} \eta = 0
%\label{phigamma}
%\end{align}
Also we trace these equations by $\partial_{A}$ and $\partial_{\mu}$ respectively. We can then write $\hat{\psi}^A$ in terms of $\lambda$ and $\eta$
%\begin{align}
%{}&\frac{3i}{8} m^4 \gamma^{\mu} \partial_{\mu} \partial_A \hat{\psi}^A + \frac{3i}{32} m^4 \gamma^{5} \partial^A \partial_{A} \lambda + \frac{3 \sqrt{6}}{32} m^6 \kappa \ \gamma^A \gamma^5 \partial_{A} \lambda - \frac{29 \sqrt{6}}{72} m^6 \kappa \ \partial_{A} \hat{\psi}^A  \nonumber \\ 
%& - \frac{\sqrt{3}}{4} m^2 \ \gamma^5 \partial^A \partial_A \eta - \frac{\sqrt{2} i}{4} m^4 \kappa \ \gamma^{A} \gamma^5 \partial_A \eta = 0
%\label{psipartial}
%\end{align}
%and
%\begin{align}
%\frac{\sqrt{3}}{3} \ \gamma^{\mu} \partial_{\mu} \partial_A \hat{{\psi}}^{A} + \frac{\sqrt{2}i}{4} m^2 \kappa \ \gamma^{\mu} \gamma^{5} \partial_{\mu} \lambda + \frac{\sqrt{6}}{6} \kappa \ \partial_{\mu} \hat{\phi}^{\mu} - \frac{\sqrt{6}}{8} \kappa \ \gamma^{\mu} \gamma^5 \partial_{\mu} \eta = 0.
%\label{phipartial}
%\end{align}
%\noindent
%From equation (\ref{psigamma}), we have 
\begin{equation}
\partial_A \hat{\psi}^A = \frac{3}{8} \gamma^5 \gamma^A \partial_A \lambda + \frac{9\sqrt{6} i}{8} m^2 \kappa \gamma^5 \lambda + 3 \sqrt{2} \kappa \gamma^5 \eta 
\label{psires}
\end{equation}
%\noindent
%Plugging this into equation (\ref{psipartial}), we get
%\begin{align}
%{}& \frac{-3i}{8} m^2 \partial_A \partial^A \lambda + \frac{17 \sqrt{6}}{12} m^4 \kappa \ \gamma^A \partial_A \lambda - \frac{87i}{4} m^6 \kappa^2 \ \lambda - {7\sqrt{2}i} m^2 \kappa \ \gamma^{A} \partial_{A} \eta \nonumber \\
%& - \frac{58\sqrt{3}}{3} m^4 \kappa^2 \ \eta - 2 \sqrt{3} \ \partial_A \partial^A \ \eta = 0.
%\label{eqdouble1}
%\end{align}
%\noindent
and a similar equation for $\partial_A \hat{\phi}^A$ 
%from equations (\ref{phigamma}) and (\ref{psires}) 
\begin{equation}
\partial_A \hat{\phi}^A = \frac{-\sqrt{3}i}{2} m^2 \gamma^5 \gamma^A \partial_A \lambda + \frac{11\sqrt{2} }{2} m^4 \kappa \ \gamma^5 \lambda - \frac{3}{4} \gamma^5 \gamma_A \partial^A \eta - \frac{7 \sqrt{6} i}{2} m^2 \kappa \gamma^5 \eta 
\end{equation}
%\noindent
%Plugging this into equation (\ref{phipartial}), we get
%\begin{align}
%\frac{-3}{8} \partial_A \partial^A \lambda - \frac{13 \sqrt{6}i}{8} m^2 \kappa \gamma^A \partial_A \lambda + \frac{11 }{2} m^4 \kappa^2 \ \lambda - {3 \sqrt{2}} \kappa \ \gamma^{A} \partial_{A} \eta - \frac{7 \sqrt{3} i}{2} m^2 \kappa^2 \ \eta = 0.
%\label{eqdouble2}
%\end{align}
\noindent
Also an equation relating $\lambda$ and $\eta$ is found. This is given by
%By combining equations (\ref{eqdouble1}) and (\ref{eqdouble2}), we get
\begin{align}
{}& -\frac{5\sqrt{6}}{24} m^4 \kappa \ \gamma^A \partial_A \lambda - \frac{109 i}{4} m^6 \kappa^2 \ \lambda - 4\sqrt{2} i\ m^2 \kappa \ \gamma^{A} \partial_{A} \eta - \frac{137 \sqrt{3}}{6} m^4 \kappa^2 \ \eta \nonumber \\
& - 2\sqrt{3} \ \partial_A \partial^A \eta = 0.
\end{align}
Upon choosing the gauge $\eta=0$
\begin{equation}
\gamma_A \partial^A \lambda + \frac{109 \sqrt{6}}{5} im^2 \kappa\ \lambda = 0
\end{equation}
This gives a Dirac type equation for the spin-1/2 helicities. 
\paragraph{}
\noindent
It should be noted that the divergence of $\hat{\phi}$ is found in terms of lambda. However, we can find a combination of $\hat{\phi}_A$ and $\hat{\psi}_A$
\begin{equation}
\hat{\phi}'_A = \hat{\phi}_A + \alpha \hat{\psi}_A
\end{equation}
such that the divergence of $\hat{\phi}'$ equals zero ($\partial^A \hat{\phi}'_A = 0$) \cite{freedman}. Then $\hat{\phi}'$ has two helicities $3/2$ and $-3/2$.

\paragraph{}
To count degrees of freedom, we are coupling supergravity to a $N=1$ supersymmetry model similar to the Wess-Zumino model. Before the coupling, supergravity contains two bosonic degrees of freedom (massless spin-$2$ graviton) and two fermionic degrees of freedom (one massless spin-$3/2$ gravitino). While the $N=1$ supersymmetry model has four spin-$0$ particles, $\varphi^A$, with only six degrees of freedom (3 times 2) since $\varphi^0$ decouples due to Fierz-Pauli choice. For this, we have six fermionic degrees of freedom forming a multiplet. Therefore, we started with an overall eight fermionic degrees of freedom and eight bosonic degrees of freedom. \par
%Counting degrees of freedom, we have before coupling to supergravity a $N=1$ supersymmetry model (similar to Wess-Zumino model) having four spin-$0$ particles. $\phi^0$ decouples due to Fierz-Pauli choice, therefore we have six (3 times 2) bosonic degrees of freedom. Also, we have six fermionic degrees of freedom forming a multiplet. This was coupled to supergravity which contains only one massless spin-2 graviton (two degrees of freedom) and one massless spin-$3/2$ gravitino (also having two degrees of freedom). Therefore, we started with an overall eight fermionic degrees of freedom and eight bosonic degrees of freedom. \\
After coupling to supergravity, we obtain $N=1$ massive representation having the same number of degrees of freedom as before coupling. The single massive spin-2 particle, with five degrees of freedom, and the single massive vector field (spin-1 particle), having three degrees of freedom, constitute the eight bosonic degrees of freedom. The fermionic degrees of freedom arise from having two massive spin-$3/2$ particles with four degrees of freedom each, and one massive vector field (spin-1 particle) having three degrees of freedom. \par
%After coupling to supergravity, we get $N=1$ massive representation containing one massive spin-2 particle with five degrees of freedom, two massive spin-$3/2$ particles with four degrees of freedom each, and one massive vector field (spin-1 particle) with three degrees of freedom. Therefore, we have as a total eight fermionic and eight bosonic degrees of freedom. \\
At the end, we are left with two massive spin-$3/2$ particles. This is similar to the $N=2$ supersymmetry in which we have two gravitinos, but there they have the same mass. However, in our case, supersymmetry is completely broken. Since it is a space-time symmetry, it is broken exactly at the same scale as the diffeomorphism breaking. Therefore, Supergravity and matter are independent and we are left with two massive spin-$3/2$ particles having completely different masses. In other words, before diffeomorphism breaking we had spin $1/2$ and not spin $3/2$, then there is no $N=2$ supersymmetry to start with. Therefore, the two gravitinos would not have the same mass. One is a genuine gravitino $\phi_{\mu}$, while the other becomes identified with a gravitino after the breaking $\psi_A$.
%It should be noted that this is similar to the $N=2$ supersymmetry in which we have two massive spin-$3/2$ particles. However, the main difference is that in $N=2$ supersymmetry we have two gravitinos with same masses. In our case, supersymmetry is completely broken where different components split and thus we are left with two massive spin-$3/2$ particles with completely different masses. 
  
\section{Conclusion}
In this paper, we gave a detailed derivation of supersymmetrizing massive gravity. Generalizing the Higgs mechanism used before to make the graviton massive, we were able to form a massive supergravity action. We started with four $N=1$ chiral superfields that break diffeomorphism invariance and local supersymmetry by the scalar component taking a vacuum expectation value. To write the full Lagrangian, we wrote the supermultiplets in terms of their component fields. First, we started by writing down all the possible D-type terms. For this we added F-type terms to satisfy all required conditions. This was coupled to supergravity using the rules of tensor calculus for chiral and vector multiplets. At the end, the degrees of freedom were analyzed and the equations of motion were obtained. In what done, we were not able to see the ghost because we were not going to the non-linear level and any emergence of such ghosts occurs at higher orders.\newline
%Starting from the formulation of massive gravity, we were able to generalize the Higgs mechanism used there to write down a massive supergravity action. To write the full Lagrangian, our analysis was carried in terms of the component fields of the supermultiplets, using the rules of tensor calculus for chiral and vector multiplets. The action was found to be composed of three D-type terms and two F-type terms. At the end, the equations of motion were analyzed.\\
Much work remains to be done as this paper shows that it is possible to construct a sensible theory of supersymmetric massive gravity with a Higgs mechanism. It remains to be seen whether it is possible to construct the action from the basic field $H_{ABC}$ by adding higher order terms. Another possibility is to generalize the simpler quadratic action \cite{quadratic} to the supersymmetric case. Also, one could analyse higher orders where ghosts may be present.

\begin{acknowledgement}
I would like to thank Professor Ali Chamseddine for suggesting the problem and for his many helpful discussions in the subject.\\
I would like to thank the American University of Beirut (Faculty of Science) for support.
\end{acknowledgement}

\appendix

\section{D type and F type terms}
\label{app1}
The products that could be formed as $D$ type terms are given by
\begin{enumerate}
\item  $H_{ABC}H_{ABC}$ where $\left(  H_{ABC}H_{ABC}\right)  ^{\ast}%
=H_{CBA}H_{CBA}=H_{ABC}H_{ABC}$ and is self adjoint.

\item $H_{ABC}H_{ACB}$ where $\left(  H_{ABC}H_{ACB}\right)  ^{\ast}%
=H_{CBA}H_{BCA}=H_{ABC}H_{BAC}$

\item $H_{ABC}H_{BCA}$ where $\left(  H_{ABC}H_{BCA}\right)  ^{\ast}%
=H_{CBA}H_{ACB}=H_{ABC}H_{BCA}$ and is self adjoint.

\item $H_{ABC}H_{CAB}$ where $\left(  H_{ABC}H_{CAB}\right)  ^{\ast}%
=H_{CBA}H_{BAC}=H_{ABC}H_{CAB}$ is self adjoint

\item $H_{ABC}H_{CBA}$ where $\left(  H_{ABC}H_{CBA}\right)  ^{\ast}%
=H_{CBA}H_{ABC}$ and is self adjoint

\item $H_{A}H_{A}^{\ast}$ is self adjoint 

\item $H_{A}H_{A}$ where $\left(  H_{A}H_{A}\right)  ^{\ast}=H_{A}^{\ast}%
H_{A}^{\ast}$

\item $\epsilon^{ABCD}H_{ABC}H_{D}$ where $\left(  \epsilon^{ABCD}H_{ABC}%
H_{D}\right)  ^{\ast}=-\epsilon^{ABCD}H_{ABC}H_{D}^{\ast}$

\item $\epsilon^{ABCD}H_{ABE}H_{CDE}$ where $\left(  \epsilon^{ABCD}%
H_{ABE}H_{CDE}\right)  ^{\ast}=\epsilon^{ABCD}H_{EBA}H_{EDC}=\epsilon
^{ABCD}H_{EAB}H_{ECD}$

\item $\epsilon^{ABCD}H_{AEB}H_{CED}$ where $\left(  \epsilon^{ABCD}%
H_{AEB}H_{CED}\right)  ^{\ast}=\epsilon^{ABCD}H_{AEB}H_{CED}$ is self adjoint

\item $\epsilon^{ABCD}H_{AEB}H_{ECD}$ where $\left(  \epsilon^{ABCD}%
H_{AEB}H_{ECD}\right)  ^{\ast}=\epsilon^{ABCD}H_{AEB}H_{CDE}$

\item $\epsilon^{ABCD}H_{EAB}H_{CDE}$ where $\left(  \epsilon^{ABCD}%
H_{EAB}H_{CDE}\right)  ^{\ast}=\epsilon^{ABCD}H_{BAE}H_{EDC}=\epsilon
^{ABCD}H_{EAB}H_{CDE}$ is self adjoint.
\end{enumerate}

%To all of this we must add $F$- type terms
%
%\begin{enumerate}
%\item $\overline{D}^{2}\left(  D\Phi_{A}\sigma^{AB}D\Phi_{B}\right)  $ 
%
%\item $\overline{D}^{2}\left(  D\Phi_{A} D\Phi^{A} \bar{D}\Phi_{B}^{*} \bar{D}\Phi^{B*} \right)  $ 
%%\item $\overline{D}^{2}\left(  D\phi_{A}D\phi_{A}\right)  $
%\end{enumerate}

\section{Notation and Convention}
\label{app2}
\allowdisplaybreaks 
metric: $g_{\mu \nu} = diag\left\{1, -1, -1 , -1\right\}$ \\
Pauli matrices: 
$\sigma^1 =
\begin{pmatrix}
0 & 1\\
1 & 0\\
\end{pmatrix}; 
\; \sigma^2 =
\begin{pmatrix}
0 & -i\\
i & 0\\
\end{pmatrix};
\; \sigma^3 =
\begin{pmatrix}
1 & 0\\
0 & -1\\
\end{pmatrix} $ \\
Dirac spinor: 
$\Psi = 
\begin{pmatrix} 
\psi_{\alpha}\\
\bar{\chi} ^ {\dot{\alpha}}
\end{pmatrix}$; adjoint Dirac spinor: $\bar{\Psi} \equiv \Psi^{*} \gamma_0 = \begin{pmatrix} \chi^{\alpha} \; \bar{\psi}_{\dot{\alpha}} \end{pmatrix}$\\
Majorana spinor 
$\Psi_M = 
\begin{pmatrix}
\psi_{\alpha} \\
\bar{\psi}_{\dot{\alpha}}
\end{pmatrix};
\quad \quad \bar{\Psi}_M = 
\begin{pmatrix}
\psi^{\alpha} \, \bar{\psi}_{\dot{\alpha}}
\end{pmatrix}$ \\
Grassman spinor: $\theta^{\alpha} = \begin{pmatrix} \theta^1 \\ \theta^2 \end{pmatrix} $; $\bar{\theta}^{\dot{\alpha}} = \begin{pmatrix} \bar{\theta}^1 \\ \bar{\theta}^2 \end{pmatrix}$ \\
Antisymmetric $\epsilon$-matrices: $\epsilon_{\alpha \beta} = \epsilon_{\dot{\alpha} \dot{\beta}} = \begin{pmatrix} 0 & 1 \\ -1 & 0 \end{pmatrix}$; $\epsilon^{\alpha \beta} = \epsilon^{\dot{\alpha} \dot{\beta}} = \begin{pmatrix} 0 & -1 \\ 1 & 0 \end{pmatrix}$ \\
$\bar{\psi}_{\dot{\alpha}} \equiv \left(\psi_{\alpha}\right)^*$ and $\chi^{\alpha} \equiv \left(\bar{\chi}^{\dot{\alpha}}\right)^*$ \\
As a convention, repeated spinor indices contracted like $^{\alpha}\,_{\alpha} \;\;\; or \;\;\; _{\dot{\alpha}}\,^{\dot{\alpha}}$
\begin{align}
\gamma^{\mu} \equiv 
\begin{pmatrix}
0 & \sigma^{\mu} \\
\bar{\sigma}^{\mu} & 0
\end{pmatrix}; \;\;\;
\gamma^5 \equiv i\gamma^0 \gamma^1 \gamma^2 \gamma^3 = 
\begin{pmatrix}
-1 & 0\\
0 & 1
\end{pmatrix}
\end{align}
There are also the following useful identities:
\begin{align}
{}& \left(\sigma^{\mu}\right)_{\alpha \dot{\beta}} = \epsilon_{\dot{\beta} \dot{\alpha}} \epsilon_{\alpha \beta} \left(\bar{\sigma}^{\mu}\right)^{\dot{\alpha} \beta} ; \;\; \left(\bar{\sigma}^{\mu}\right)^{\dot{\alpha} \beta} = \epsilon^{\beta \alpha} \epsilon^{\dot{\alpha} \dot{\beta}} \left(\sigma^{\mu}\right)_{\alpha \dot{\beta}} \nonumber \\
& \left(\sigma^{\mu}\right)_{\alpha \dot{\alpha}} \left(\bar{\sigma}^{\nu}\right)^{\dot{\alpha} \alpha} = Tr \left(\sigma^{\mu} \bar{\sigma}^{\nu}\right) = 2\eta^{\mu \nu} \nonumber \\
& \left(\sigma^{\mu}\right)_{\alpha \dot{\alpha}} \left(\sigma_{\mu}\right)_{\beta \dot{\beta}} = 2\epsilon_{\alpha \beta} \epsilon_{\dot{\alpha} \dot{\beta}} \nonumber \\
& \left(\sigma^{\mu}\right)_{\alpha \dot{\alpha}} \left(\bar{\sigma}_{\mu}\right)^{\beta \dot{\beta}} = 2\delta_{\alpha}^{\beta} \delta_{\dot{\alpha}}^{ \dot{\beta}} \nonumber \\
& \left(\sigma^{\mu \nu}\right)_{\alpha}\,^{\beta} \equiv \frac{i}{4} \left(\sigma^{\mu} \bar{\sigma}^{\nu} - \sigma^{\nu} \bar{\sigma}^{\mu}\right)_{\alpha}\,^{\beta}; \;\;\; \left(\bar{\sigma}^{\mu \nu}\right)^{\dot{\alpha}}\,_{\dot{\beta}} \equiv \frac{i}{4} \left(\bar{\sigma}^{\mu} {\sigma}^{\nu} - \bar{\sigma}^{\nu} {\sigma}^{\mu}\right)^{\dot{\alpha}}\,_{\dot{\beta}}\, \nonumber \\
& \sigma^{\mu \nu} = \frac{1}{2i} \epsilon^{\mu \nu \rho \sigma} \sigma_{\rho \sigma}; \;\;\; \bar{\sigma}^{\mu \nu} = \frac{-1}{2i} \epsilon^{\mu \nu \rho \sigma} \bar{\sigma}_{\rho \sigma} \nonumber \\
& \left(\sigma^{\mu} \bar{\sigma}^{\nu} + \sigma^{\nu} \bar{\sigma}^{\mu}\right)_{\alpha}\,^{\beta} = 2 \eta^{\mu \nu} \delta_{\alpha}^{\beta}\,; \;\;\; \left(\bar{\sigma}^{\mu} {\sigma}^{\nu} + \bar{\sigma}^{\nu} {\sigma}^{\mu}\right)^{\dot{\alpha}}\,_{\dot{\beta}} = 2 \eta^{\mu \nu} \delta^{\dot{\alpha}}_{\dot{\beta}} \nonumber \\
& \sigma^{\mu} \bar{\sigma}^{\nu} \sigma^{\rho} + \sigma^{\rho} \bar{\sigma}^{\nu} \sigma^{\mu} = 2 \left(\eta^{\mu \nu} \sigma^{\rho} + \eta^{\nu \rho} \sigma^{\mu} - \eta^{\mu \rho} \sigma^{\nu}\right) \nonumber \\
& \bar{\sigma}^{\mu} {\sigma}^{\nu} \bar{\sigma}^{\rho} + \bar{\sigma}^{\rho} {\sigma}^{\nu} \bar{\sigma}^{\mu} = 2 \left(\eta^{\mu \nu} \bar{\sigma}^{\rho} + \eta^{\nu \rho} \bar{\sigma}^{\mu} - \eta^{\mu \rho} \bar{\sigma}^{\nu}\right) \nonumber \\
& \sigma^{\mu} \bar{\sigma}^{\nu} \sigma^{\rho} - \sigma^{\rho} \bar{\sigma}^{\nu} \sigma^{\mu} = -2i \epsilon^{\mu \nu \rho \kappa} \sigma_{\kappa} \nonumber \\
& \bar{\sigma}^{\mu} {\sigma}^{\nu} \bar{\sigma}^{\rho} - \bar{\sigma}^{\rho} {\sigma}^{\nu} \bar{\sigma}^{\mu} = 2i \epsilon^{\mu \nu \rho \kappa} \bar{\sigma}_{\kappa} \nonumber \\
& Tr\left(\sigma^{\mu} \bar{\sigma}^{\nu} \sigma^{\rho} \bar{\sigma}^{\kappa}\right) = 2\left(\eta^{\mu\nu} \eta^{\rho\kappa} + \eta^{\mu\kappa} \eta^{\nu\rho} - \eta^{\mu\rho} \eta^{\nu\kappa}  - i\epsilon^{\mu \nu \rho \kappa} \right)
\label{identities}
\end{align}
where $\epsilon_{0123} = +1$. 
\begin{align*}
\theta^{\alpha} \theta^{\beta} = - \frac{1}{2} \epsilon^{\alpha \beta} \left(\theta \theta\right); \;\;\; \bar{\theta}^{\dot{\alpha}} \bar{\theta}^{\dot{\beta}} = \frac{1}{2} \epsilon^{\dot{\alpha} \dot{\beta}} \left(\bar{\theta} \bar{\theta}\right); \nonumber \\
\theta_{\alpha} \theta_{\beta} =  \frac{1}{2} \epsilon_{\alpha \beta} \left(\theta \theta\right); \;\;\; \bar{\theta}_{\dot{\alpha}} \bar{\theta}_{\dot{\beta}} = -\frac{1}{2} \epsilon^{\dot{\alpha} \dot{\beta}} \left(\bar{\theta} \bar{\theta}\right)
\end{align*}
The derivatives with respect to a Grassmann variable are defined as follows:
\begin{equation*}
\partial_{\alpha} \equiv \frac{\partial}{\partial \theta^{\alpha}}; \;\; \partial^{\alpha} \equiv -\epsilon^{\alpha \beta} \partial_{\beta}; \;\; \bar{\partial}_{\dot{\alpha}} \equiv \frac{\partial}{\partial \bar{\theta}^{\dot{\alpha}}}; \;\; \bar{\partial}^{\dot{\alpha}} \equiv -\epsilon^{\dot{\alpha} \dot{\beta}} \bar{\partial}_{\dot{\beta}}.
\end{equation*}
This implies that
\begin{align*}
{}& \partial_{\alpha} \theta^2 = 2 \theta_{\alpha}; \;\; \partial^{\alpha} \theta^2 = -2 \theta^{\alpha};\nonumber \\
& \bar{\partial}_{\dot{\alpha}} \bar{\theta}^2 = -2 \bar{\theta}_{\dot{\alpha}}; \;\; \bar{\partial}^{\dot{\alpha}} \bar{\theta}^2 = 2 \bar{\theta}^{\dot{\alpha}}.
\end{align*}
\begin{align*}
{}& D_{\alpha} = \frac{\partial}{\partial \theta^{\alpha}} + i \sigma^{\mu}_{\alpha \dot{\alpha}} \bar{\theta}^{\dot{\alpha}} \partial_{\mu} \\
& \bar{D}_{\dot{\alpha}} = -\frac{\partial}{\partial \bar{\theta}^{\dot{\alpha}}} - i \theta^{\alpha} \sigma^{\mu}_{\alpha \dot{\alpha}} \partial_{\mu} 
\end{align*}

\end{document}